\documentclass[center,twocolumn,a4paper,bold]{paper}

\pdfoutput=1

\newcommand{\cpponly}[1]{}
\newcommand{\arxivonly}[1]{#1}
\newcommand{\cpporarxiv}[2]{#2}

\arxivonly{
  \usepackage[hmargin=0.75in,vmargin={1.5in,1.2in},columnsep=0.2in]{geometry}
}

\usepackage[utf8]{inputenc}

\arxivonly{
  \usepackage[T1]{fontenc}
  \usepackage{microtype}
  \usepackage[osf]{libertine}
  \usepackage[scaled=0.83]{beramono}
}

\arxivonly{
  \titlefont{\huge\bfseries}
  \subtitlefont{\large \bfseries}
  \def\abstract{\subsection*{\abstractname}}
}

\usepackage{comment}
\usepackage{url}
\usepackage{amssymb}
\usepackage{xifthen} 
\usepackage{mathtools} 
\usepackage{balance} 

\arxivonly{
  \usepackage[style=authoryear,backend=bibtex]{biblatex}
  \addbibresource{hott.bib}

  \let\citet\textcite
  \let\citep\parencite
}
\newcommand{\sortas}[1]{}

\cpporarxiv{
  \usepackage{color}
}{
  \usepackage{xcolor}
}

\definecolor{dkblue}{rgb}{0,0.1,0.5}
\definecolor{lightblue}{rgb}{0,0.5,0.5}
\definecolor{dkgreen}{rgb}{0,0.4,0}
\definecolor{dk2green}{rgb}{0.4,0,0}
\definecolor{dkviolet}{rgb}{0.6,0,0.8}
\definecolor{darkblue}{rgb}{0.0,0.0,0.6}
\definecolor{grey}{rgb}{0.5,0.5,0.5}

\cpporarxiv{
  \usepackage[
            unicode]{hyperref}
            \def\UrlBreaks{\do\/\do-\do_}
}{
  \usepackage[
            colorlinks,
            linkcolor=darkblue,
            citecolor=darkblue,
            urlcolor=darkblue,
            unicode]{hyperref}
            \csname g@addto@macro\endcsname{\UrlBreaks}{\do-\do_}
}

\newcommand{\arxiv}[1]{\href{http://arxiv.org/abs/#1}{\nolinkurl{arXiv:#1}}}

\usepackage{listings}

\newcommand{\type}{\mathsf{Type}}
\newcommand{\prop}{\mathsf{Prop}}

\newcommand{\hprop}{\mathsf{hProp}}

\begin{document}

\cpponly{ 
  \toappear
}

\setlength{\pdfpageheight}{\paperheight}
\setlength{\pdfpagewidth}{\paperwidth}

\cpporarxiv{
  \title{The HoTT Library}
  \subtitle{A Formalization of Homotopy Type Theory in Coq}
}{
  \title{The HoTT Library}
  \subtitle{A formalization of homotopy type theory in Coq\medskip}
}

\newcommand{\maybethanks}[1]{\ifthenelse{\equal{#1}{}}{}{\thanks{#1}}}
\newcommand{\ourauthorinfo}[4]{\cpporarxiv
  {\authorinfo{#1\maybethanks{#2}}{#3}{#4}}
  {#1\maybethanks{#2} \\ #3 \\ \nolinkurl{#4}}%
}
\newcommand{\firstauthorinfo}[4]{\ourauthorinfo{#1}{#2}{#3}{#4}}
\newcommand{\nextauthorinfo}[4]{\cpporarxiv
  {\ourauthorinfo{#1}{#2}{#3}{#4}}
  {\and \ourauthorinfo{#1}{#2}{#3}{#4}}
}
\newcommand{\authors}[1]{\cpporarxiv
  {#1}
  {\author{#1}}
}

\cpporarxiv{
  \newcommand{\fixouterwidth}[1]{#1}
  \newcommand{\fixmidwidth}[1]{#1}
}{
  \newlength\stextwidth
  \newcommand\makesamewidth[3][c]{%
    \settowidth{\stextwidth}{#2}%
    \makebox[\stextwidth][#1]{#3}%
  }
  
  \newcommand{\widestouter}{University of Ljubljana, Slovenia}
  \newcommand{\widestmid}{\nolinkurl{mattam@mattam.org}}
  \newcommand{\fixouterwidth}[1]{\makesamewidth{\widestouter}{#1}}
  \newcommand{\fixmidwidth}[1]{\makesamewidth{\widestmid}{#1}}
}

\authors{
\firstauthorinfo{Andrej Bauer}
           {This material is based upon work supported by the Air Force Office of Scientific Research, Air Force Materiel Command, USAF under Award No. FA9550-14-1-0096}
           {\fixouterwidth{University of Ljubljana, Slovenia}}
           {Andrej.Bauer@andrej.com}%
\nextauthorinfo{Jason Gross}
           {}
           {\fixmidwidth{MIT, USA}}
           {jgross@mit.edu}%
\nextauthorinfo{Peter LeFanu Lumsdaine}
           {This material is based upon work supported by the National Science Foundation under Grant No.~DMS-1128155.}
           {\fixouterwidth{Stockholm University, Sweden}}
           {p.l.lumsdaine@math.su.se}%
\nextauthorinfo{Michael Shulman}
           {This material is based on research sponsored by The United States Air Force Research Laboratory under agreement number FA9550-15-1-0053.  The U.S.~Government is authorized to reproduce and distribute reprints for Governmental purposes notwithstanding any copyright notation thereon.  The views and conclusions contained herein are those of the authors and should not be interpreted as necessarily representing the official policies or endorsements, either expressed or implied, of the United States Air Force Research Laboratory, the U.S.~Government, or Carnegie Mellon University.}
           {\fixouterwidth{University of San Diego, USA}}
           {shulman@sandiego.edu}%
\nextauthorinfo{Matthieu Sozeau}
           {This work has been partly funded by the CoqHoTT ERC Grant 637339.}
           {Inria, France}
           {mattam@mattam.org}%
\nextauthorinfo{Bas Spitters}
           {This research was partially supported by the Guarded Homotopy Type Theory project, funded by the Villum Foundation, project number 12386.}
           {\fixouterwidth{Aarhus University, Denmark}}
           {spitters@cs.au.dk}%
}

\maketitle

\newcommand{\ourabstract}[1]%
  {\cpporarxiv{\abstract{#1}}{ \begin{abstract}#1\end{abstract}}}

\ourabstract{
  We report on the development of the \emph{HoTT library}, a formalization of homotopy
  type theory in the Coq proof assistant. It formalizes most of basic homotopy type
  theory, including univalence, higher inductive types, and significant amounts of
  synthetic homotopy theory, as well as category theory and modalities. The library
  has been used as a basis for several independent developments. We discuss the decisions
  that led to the design of the library, and we comment on the interaction of homotopy
  type theory with recently introduced features of Coq, such as universe polymorphism
  and private inductive types.
}

\cpponly{


}


\section{Introduction}

Homotopy type theory is a novel approach to developing mathematics in Martin-L\"of’s type theory, based on interpretations of the theory into abstract
homotopy-theoretic settings such as certain higher toposes \citep{kapulkin2012simplicial,shulman2012inverse}.
The connection between type theory and homotopy theory is originally due to \citep{AW} and \citep{VV}.

Identity types are interpreted as path spaces, and type equivalence as homotopy
equivalence. Type-theoretic constructions correspond to homotopy-invariant
constructions on homotopy types. In addition, homotopical intuition gives rise to entirely new
type-theoretic notions, such as higher inductive types and Voevodsky's univalence axiom.
One can even develop homotopy theory in the language of type theory in a ``synthetic''
manner, treating (homotopy) types as a primitive notion.

The first formalization of homotopy type theory in a proof assistant was Voevodsky's \emph{Foundations} library
implemented in Coq, now called the \emph{UniMath} project~\citep{UniMath}. Here we present the second major such library, \emph{the HoTT library},
also implemented in Coq, with somewhat different goals from
those of UniMath. The library is freely available.\footnote{\url{http://github.com/HoTT/HoTT} or
the Coq OPAM package manager} 

Coq word count reports that the library contains 16800 lines of specifications, 13000
lines of proofs, and 4500 lines of comments. The library is self-sufficient, completely
replacing the Coq standard library (which is incompatible with homotopy type theory) with a
bare minimum necessary for basic Coq tactics to function properly (see~\S\ref{sec:coq-changes}).

\paragraph{Contributions}
The HoTT library provides a substantive formalization of homotopy type theory. It
demonstrates that univalent foundations (cf.~\S\ref{Overview}) provide a workable setup
for formalization of mathematics. The library relies on advanced features of Coq
(cf.~\S\ref{sec:coq-features}), such as automatic handling of universe polymorphism
(cf.~\S\ref{univpoly}) and type classes (cf.~\S\ref{sec:type-classes}), management of
opaque and transparent definitions (cf.~\S\ref{sec:opaque}), and automation
(cf.~\S\ref{sec:automation}). We used private inductive types to implement higher
inductive types (cf.~\S\ref{HIT}), and the Coq module system to formalize modalities
(cf.~\S\ref{sec:modules}). Our development pushed Coq's abilities, which prompted the
developers to extend and modify it for our needs (cf.~\S\ref{sec:coq-changes}), and to
remove several bugs, for which we are most thankful. Overall, the success of the project
relies on careful policies and software-engineering approaches that keep the library
maintainable and usable (cf.~\S\ref{sec:software-engineering}). We relate our work to
other extensive implementations of homotopy type theory in~\S\ref{sec:related-work}.

\paragraph{Consistency}

A major concern for any piece of formalized mathematics is trust. The
HoTT library uses much more than just Martin-L\"of type theory,
including the univalence axiom, pattern matching, universe
polymorphism, type classes, private inductive types, second-class
modules, and so on; how do we know these form a consistent system?
This is a major concern not just for us, but for every user of complex
proof assistants, and has been addressed many times in the past.

In Coq, the typechecker’s kernel is the final gatekeeper. This is fairly large, in part due to
the inclusion of the module system. However, other advanced features
such as type classes and tactics are outside the kernel
and hence do not endanger consistency. So, provided we trust the
kernel, the remaining questions are the consistency of our
axioms, universe-polymorphic modules, and the implementation of private
inductive types.

There are several possible ways to tackle these questions.  So far, the primary
method available for homotopy type theory is semantic: constructing a
model of the theory in some other trusted theory (such as ZFC).  While
this has been done for various fragments of the theory,
combining them all to give a unified semantic account of homotopy type
theory together with all the features in Coq's kernel seems a daunting task.

For this reason, UniMath avoids almost all of Coq’s features (even e.g.~record types),
restricting itself as far as possible to standard Martin-L\"of
type theory (except for assuming
$\type\,{:}\,\type$ throughout, to simulate Voevodsky's resizing rules). However, this restriction cannot be
enforced by the kernel.
We feel rather that proof assistants and computerized formalization of mathematics are at such an
early stage that it is well worth experimenting, even at the risk of introducing an inconsistency
(which is fairly slight, due to the known semantic accounts of fragments of the theory).
In any case, the skeptical reader
should keep in mind that the standard of rigor in formalized proofs is at least a great deal
higher than the generally accepted level of rigor in traditional written mathematics.


\section{Basics}\label{Overview}


We assume basic familiarity with homotopy type theory~\citep{hottbook}, and with the Coq
proof assistant~\citep{Coq}.
There is a large overlap between the contents of the HoTT library and the contents of the
book~\citep{hottbook}, which we refer to as the ``HoTT book''. The library provides an
automatically generated file linking the constructions in the book with the corresponding Coq
code.\footnote{\url{https://hott.github.io/HoTT/coqdoc-html/HoTTBook.html}}

\paragraph{Basic type formers and their identity types}

The core of the library is in the \texttt{Basics} and \texttt{Types}
directories.  These files formalize fundamental results about the higher
groupoid structure of identity types, and the identity types of basic types and type formers, such
as the empty and unit types, universes, $+$, $\times$, $\Pi$, and $\Sigma$.  This covers most of
Chapter~2 of the HoTT book, as well as parts of Chapters~3 and~7
(basic definitions and facts about $n$-types), Chapter~4 (equivalences; see below), and Chapter~5 (basic facts
about W-types).

The \texttt{Basics} directory contains absolutely basic facts
applicable to all types; while the \texttt{Types} directory is
organized with one file for each standard type former, roughly
matching the sections in Chapter~2 of the HoTT book.  Some other
basic facts from the first part of the HoTT book can be found in the
root \texttt{theories} directory, such as the comparison of different
definitions of equivalence (see below) and the proof that univalence
implies function extensionality.

\paragraph{Equivalences}

The HoTT book devotes most of Chapter~4 to discussing
various notions of equivalence. After showing that a large class of
them are equivalent, in a precise way, one can be agnostic on paper
about which is meant.  However, for a formalization we
need to choose a particular definition.

The intuitive notion of isomorphism or homotopy equivalence consists
of $f : A \to B$ and $g : B \to A$ which are inverses of each other,
up to homotopy.  However, in homotopy type theory the type of pairs
$f,g$ equipped with two such homotopies is ill-behaved, so one needs
to refine it somehow.

We have chosen to use the notion called a \emph{half-adjoint
  equivalence} in the HoTT book, which adds to this type a single
coherence condition between the two homotopies.  (The condition is one
of the triangle identities involved in an ``adjoint equivalence'' in
category theory; the other one is then provable, but should not be
assumed as data or the homotopy type would be wrong again.)  Since
this is ``the'' notion of equivalence in the library, we call it
simply an \emph{equivalence}.  Other possible options are Voevodsky's
definition of an equivalence as a map whose homotopy fibers are
contractible, or Joyal's suggestion of a map equipped with
separate left and right homotopy inverses.  We do prove the
equivalence of all these definitions (in
\texttt{EquivalenceVarieties}).

However, we believe that half-adjoint equivalences are a better choice
for the standard notion of equivalence in a formalization.  This is because
the most common way to construct an equivalence, and to use an
equivalence, is by way of the ``incoherent'' notion consisting of two
functions and two homotopies, called a \emph{quasi-inverse}
in the HoTT book, and half-adjoint equivalences record all
this data.  That is, usually we construct an equivalence
by exhibiting its homotopy inverse, and then apply a
``coherentification'' result.  With half-adjoint equivalences
represented as a Coq record, all the data of a
quasi-inverse (plus the extra coherence) is stored exactly as supplied when an
equivalence is defined.  This applies in particular to the homotopy
inverse, but also to the witnessing homotopies, though the
``coherentification'' process alters one of these homotopies, so if we
want to preserve them both we have to manually prove the extra
coherence property.

In addition to the definitions of equivalence appearing in the HoTT
Book, we also consider two others.  One is a ``relational
equivalence'' (a relation under which each element of either type is related to a unique element of the other)
which has the advantage of being judgmentally
invertible, though it increases the universe level:
\begin{lstlisting}
Record RelEquiv A B :=
  { equiv_rel : A -> B -> Type;
    rcf : forall a, Contr { b : B & equiv_rel a b };
    rcg : forall b, Contr { a : A & equiv_rel a b } }.
\end{lstlisting}
The other involves ``$n$-path-splitness'', which says that the induced
maps on the first $n$ path spaces are split surjections.
\begin{lstlisting}
Fixpoint PathSplit (n : nat) `(f : A -> B) : Type
  := match n with
       | 0 => Unit
       | S n => (forall a, hfiber f a) *
                forall x y, PathSplit n (@ap _ _ f x y)
     end.
\end{lstlisting}
For $n>1$, this is equivalent to being an equivalence.  This
definition has the advantage that when $A$ and $B$ are function-types
and $f$ is a precomposition map, we can reformulate it to use
homotopies rather than equalities, yielding a notion of
``precomposition equivalence'' (called \lstinline|ExtendableAlong|)
that often avoids function extensionality.  This is particularly
useful for the universal property of modalities (\S\ref{sec:modules}).

\paragraph{Finite sets}
We define standard finite types \lstinline|Fin n| as usual,
\begin{lstlisting}
Fixpoint Fin (n : nat) : Type :=
  match n with
     | 0 => Empty
     | S n => Fin n + Unit
   end.
\end{lstlisting}
and then finite types as those that are merely\footnote{As in the HoTT book, ``merely'' signifies the use of a propositional truncation, in this instance an existence of an equivalence rather
than a concretely given one.} equivalent to the standard ones:
\begin{lstlisting}
Class Finite (X : Type) :=
  { fcard : nat ;
    merely_equiv_fin : merely (X <~> Fin fcard) }.
\end{lstlisting}
Perhaps surprisingly, being finite is still a mere proposition, because
a set is isomorphic to at most one canonical finite set.  Thus, we could
have truncated the dependent sum and gotten an equivalent definition, but it
would be less convenient to reason about.

\paragraph{Pointed types}
\label{sec:pointed-types}

We provide a general theory of pointed types. The theory is
facilitated by a tactic which often allows us to pretend that pointed maps and homotopies preserve basepoints \emph{strictly}.  We have carefully defined pointed maps \lstinline|pMap| and pointed homotopies \lstinline|pHomotopy| so that when destructed, their second components are paths with right endpoints free, to which we can apply Paulin-Mohring path-induction. The theory of pointed types uses type
classes, since the base point can usually be found automatically:
\begin{lstlisting}
Class IsPointed (A : Type) := point : A.
Record pType :=
  { pointed_type : Type ;
    ispointed_type : IsPointed pointed_type }.
Coercion pointed_type : pType >-> Sortclass.
Record pMap (A B : pType) :=
 { pointed_fun : A -> B ;
   point_eq : pointed_fun (point A) = point B }.
Record pHomotopy {A B : pType} (f g : pMap A B) :=
 { pointed_htpy: f == g ;
   point_htpy: 
     pointed_htpy (point A) @ point_eq g = 
     point_eq f }.
\end{lstlisting}
Here \lstinline|@| denotes the concatenation of paths.

\paragraph{Category theory}

The library also includes a large development of category theory,
following Chapter~9 of the HoTT book and~\citep{ahrens2015univalent}. This part of the library was presented in detail in~\citep{grosscat}, from which we
quote only the following:
\begin{quote}
  We wound up adopting the Coq
  version under development by homotopy type theorists, making critical use
  of its stronger universe polymorphism and higher inductive
  types\dots [which] can simplify the Coq user experience
  dramatically\dots
\end{quote}
The category theory library employs a different style of formalization
from the core library, using so-called ``blast'' tactics that
automatically try many lemmas to produce a proof by brute
force.  We avoid this approach in the core library to make proofs more
readable and give better control over proof terms, cf.~\S\ref{sec:opaque}.

\citet{timany-jacobs:category-theory-extended}\footnote{\url{https://github.com/amintimany/Categories-HoTT}} provide another
extensive library for category theory over the HoTT library.

\paragraph{Synthetic Homotopy theory}

The library also contains a variety of other
definitions and results, many relevant to
synthetic homotopy theory or higher category theory.  This includes
classifying spaces of automorphism groups, the Cantor
space, the theory of idempotents with the results of~\citep{shulman2016idempotents}, and the definition
of $\infty$-groups with actions. Such non-trivial additions to the core
provide strong evidence that the overall design is sustainable and usable.


\section{Features of Coq}
\label{sec:coq-features}

In our library we made use of a number of useful features of Coq, some
established and others relatively new.  In this section we discuss
most of these; two of the most substantial we give their own sections
(\S\ref{HIT} and \S\ref{sec:modules}).

\subsection{Universe polymorphism}\label{univpoly}

Coq's type theory supports an infinite progression of cumulative
universes, and constructions that are polymorphic in the universes~\citep{sozeau2014universe}. Our library heavily relies
on the flexible treatment of universe polymorphism, which allows us to forego
universe annotations in specifications. However, one needs to be careful
because Coq's heuristic for minimizing the number of universes and constraints
occasionally produces undesirable constraints, or just does not cope with them very well.
For instance, the formalization of cumulative hierarchy of sets by \citet{Ledent} induced
unexpected universe inconsistencies that could only be resolved with
explicit universe constraint annotations. These were indeed implemented
in response to our troubles, and are now part of the standard Coq
distribution.



\citet{KULeuven-513812} propose cumulative inductive
types in Coq. It seems this would indeed alleviate most of the
issues in our library, too.

\paragraph{Object classifier}

Following Section~4.8 of the HoTT book, we formalized the proof that the
universe is an object classifier, a fundamental concept in higher
topos theory.
From this we can show that the propositions \lstinline|hProp| form a \emph{large}
subobject classifier.  In fact, we show more generally that for each
$P:\type\to\type$ taking values in mere
propositions, we have a classifier for maps whose fibers satisfy $\mathtt{P}$; this also
includes all modalities (see \S\ref{sec:modules}).

In this part of the library, universe polymorphism works nicely and leads to
a satisfactory treatment of subobject classifiers in a predicative setting.
As a test, we verified that in \lstinline|hSets| epimorphisms are
surjective. The proof looks like the usual impredicative one, but
in fact, is entirely predicative thanks to universe polymorphism.


\subsection{Type classes}\label{sec:type-classes}

Coq's type class system is a very convenient mechanism for automatic derivation of
instances of structures. The instance search follows a logic programming discipline which
is sometimes difficult to predict and control, a potential problem for proof-relevant
settings such as homotopy type theory. Nevertheless, it is safe to use the type class
mechanism as long as it is employed to find only mere propositions, as all their instances are
equal. Luckily, being an equivalence, deriving a truncation level, and placing a type into
a subuniverse, are all of this kind, and so our library does use type classes in these
non-problematic cases. We follow the development style of the math-classes
library~\citep{math-classes,KrebbersSpitters,casteran:hal-00702455}, but since we have
a better behaved equality and quotients, we can avoid an extensive use of setoids.

As is well known, one can only push the type class mechanism so far. For
instance, we cannot add a rule for automatic instantiation of inverses of equivalence, as
it makes Coq look for ``the inverse of the inverse of the inverse \dots''.

\paragraph{Transfer from $\Sigma$-types to record types}

Type classes are (dependently typed) record types, so many of the central concepts in the
library are expressed as record types (rather than nested $\Sigma$-types). To avoid proving a
series of general lemmas over and over, separately for each record type,
we implemented a tactic \lstinline|issig| that allows us to automatically transfer general facts about
iterated $\Sigma$-types to record types. For example, \lstinline|Contr| and
\lstinline|IsEquiv| are record types, and the tactic is able to automatically prove that
\begin{lstlisting}
  forall A,  {x : A & forall y:A, x = y} <~> Contr A.
\end{lstlisting}
and
\begin{lstlisting}
  forall A B, forall f : A -> B,
   IsEquiv f <~>
     {g : B -> A &
       {r : Sect g f &
       {s : Sect f g &
         forall x : A, r (f x) = ap f (s x) }}}.
\end{lstlisting}
%
A fairly substantial use of this tactic arises in the development of
factorization systems (e.g.\ epi-mono factorization), for the proof of uniqueness of factorizations (our \lstinline|path_factorization|, Theorem
7.6.6 in the HoTT book).   The book proof expresses the uniqueness using a threefold nesting of
$\Sigma$-types followed by a threefold cartesian product. In the formalization this
corresponds to a record type with six fields. The formal proof requires delicate
transformations between records and $\Sigma$-types that would be very laborious to perform
without \lstinline|issig|.

\paragraph{The $\eta$-rule for record types}

Recent versions of Coq support an extensionality $\eta$-rule for record types,
implemented by primitive projections.
When we adapted the library to use primitive projections, the compilation time for the
entire library dropped by a factor of two. We were also able to replace eight different
kinds of natural transformations with a single one, remove applications of function
extensionality from a number of proofs, and greatly speed up the tactics for transfer
between $\Sigma$-types are record types.

\paragraph{Type classes for axioms}

We also use type classes to track the use of the univalence and
function extensionality axioms.  These axioms are defined to depend on
an instance of a class that inhabits a dummy type.  For instance, here
is function extensionality:
\begin{lstlisting}
Monomorphic Axiom dummy_fe : Type0.
Monomorphic Class Funext := { fev : dummy_fe }.
Axiom isequiv_apD10 :
  forall `{Funext} (A : Type) (P : A -> Type) f g,
    IsEquiv (@apD10 A P f g).
\end{lstlisting}
Here, \lstinline|apD10| is the canonical map from paths in the function type to homotopies between functions.

Any theorem that uses function extensionality must then state this
fact explicitly with a \lstinline|`{Funext}| assumption.  This makes
it easy to tell which parts of the library depend on which axioms,
without the need for \lstinline|Print Assumptions|.  We make
\lstinline|Funext| inhabit a dummy type rather than the actual type of
\lstinline|isequiv_apD10| so that the latter can be used at multiple
universe levels with only a single assumption of \lstinline|Funext|.


\subsection{Transparency and Opacity}\label{sec:opaque}
Since equality in homotopy type theory is proof-relevant, our lemmas are more often transparent than in most Coq
developments. This means that much more care must be taken to construct the
right, coherent, proof terms. An example is the proof of
the Eckmann-Hilton theorem for identity types (Theorem 2.1.6 of the HoTT book), which is given explicitly using (higher) composites and whiskerings:
\begin{lstlisting}
Definition eckmann_hilton
  {A : Type} {x:A} (p q : 1 = 1 :> (x = x)) :
  p @ q = q @ p
  :=  (whiskerR_p1 p @@ whiskerL_1p q)^
      @ (concat_p1 _ @@ concat_p1 _)
      @ (concat_1p _ @@ concat_1p _)
      @ (concat_whisker _ _ _ _ p q)
      @ (concat_1p _ @@ concat_1p _)^
      @ (concat_p1 _ @@ concat_p1 _)^
      @ (whiskerL_1p q @@ whiskerR_p1 p).
\end{lstlisting}
Here \lstinline|@| denotes concatenation of paths and \lstinline|@@|
denotes horizontal composition of 2-dimensional paths.
It might be easier to find \emph{a} proof using rewrite-type tactics, but we want this particular one.

We could benefit from good support for construction of explicit proof terms.
Sozeau's dependent pattern matching compiler~\citep{sozeau2010equations} aims to
provide such support in Coq. It is similar to how Agda works. Unfortunately, it currently depends on both the
equality type being in $\prop$ and uniqueness of identity proofs,
which is incompatible with univalence. This issue has been solved in
Agda~\citep{cockx2014pattern} and is currently being adapted to Coq as
part of the Equations package~\citep{mangin2015equations}.
The prelude of the HoTT library can already be processed using Equations, and we hope
to be able to use it for the entire library in the future.

\subsection{Automation}\label{sec:automation}

Once we collected a large number of lemmas about paths, we organized
them in a rewrite database and used it to simplify proofs.
Currently, we rewrite terms in a standard fashion, that is with
$J$-elimination. However, we are likely to obtain better proof terms
by using the technology of generalized rewriting~\citep{sozeau2010new}, which would also
know how to rewrite with equivalences, and other suitable general relations.
For instance,
\begin{lstlisting}
forall {A B:Type},
  IsHProp A -> (A <~>B) -> IsHProp B
\end{lstlisting}
can be proved by rewriting without invoking
the univalence axiom.
This is done by first showing that the basic type constructors $\Pi,\Sigma$
respect equivalence, and that contractibility transfers along
equivalence. All concepts built from these will then also respect
equivalence. So, surprisingly, even now that we have quotients in our
type theory, the technology initially developed for setoid rewriting
is still useful.

Two other tactics are worth mentioning.
First, we have a tactic \lstinline|transparent assert|, like ordinary \lstinline|assert| (allowing on-the-fly interactive proof of assertions) except that the term produced remains transparent in the rest of the proof. (Meanwhile, this has been ported for inclusion in standard Coq.)
Second, we implemented a custom version of the \lstinline|apply| tactic which uses a more powerful unification algorithm than the one used by the standard \lstinline|apply| tactic. This idea is inspired by similar techniques in ssreflect~\citep{gonthier:inria-00258384}.



\section{\cpporarxiv{Higher Inductive Types}{Higher inductive types}}\label{HIT}

Higher inductive types are one of the main novelties in homotopy type
theory. Where usual inductive types allow us to freely generate terms
in a type, \emph{higher} inductive types also allow us to
\emph{freely} generate equalities. Examples include the interval (two
points and a path), the circle (a point and a self loop), suspensions,
set quotients, and more complex examples that we shall discuss briefly.

Coq does not implement higher inductive types natively, so we simulate
them using Licata's trick~\citep{licata2013calculating}.  This method
was originally used in Agda; to make it possible in Coq, experimental
\emph{private inductive types} had to be added~\citep{Bertot}.  Private
inductive types are defined inside a module, within which they behave
as usual.  Outside the module their induction principles and pattern
matching are no longer available; but functions that were defined
inside the module using those principles can still be called, and
importantly still compute on constructors.

Licata's trick uses this to implement higher inductive types whose
induction principles compute definitionally on point-constructors.
For instance, the definition of the interval is shown in
Figure~\ref{fig:interval}.  Here \lstinline|#| denotes transport, \lstinline|@| is concatenation of paths, and \lstinline|apD| is application of a dependent function
to a path.  Importantly, \lstinline|interval_rec a b p zero|
\emph{reduces} to \lstinline|a|, because \lstinline|zero| is actually
a constructor of \lstinline|interval|.
The axiom \lstinline|seg| could be used to derive a
contradiction within the module; but this is not possible
outside the module, so all we need to do is check that the code inside
the module is safe.

\begin{figure}[t]
\centering
\begin{lstlisting}
Module Export Interval.

Private Inductive interval : Type1 :=
  | zero : interval
  | one : interval.

Axiom seg : zero = one.

Definition interval_ind (P : interval -> Type)
  (a : P zero) (b : P one) (p : seg # a = b)
  : forall x:interval, P x
  := fun x => (match x return _ -> P x with
                | zero => fun _ => a
                | one  => fun _ => b
              end) p.

Axiom interval_ind_beta_seg :
  forall (P : interval -> Type)
    (a : P zero) (b : P one) (p : seg # a = b),
    apD (interval_ind P a b p) seg = p.

End Interval.

Definition interval_rec (P : Type) (a b : P) (p : a = b)
  : interval -> P :=
  interval_ind (fun _ => P) a b (transport_const _ _ @ p).

Definition interval_rec_beta_seg
  (P : Type) (a b : P) (p : a = b) :
  ap (interval_rec P a b p) seg = p.
\end{lstlisting}
  \caption{The higher inductive interval}
  \label{fig:interval}
\end{figure}

Note that in \lstinline|interval_ind| rather than simply matching
on $x$, we apply the match statement to the path hypothesis $p$ (but
then never use it).  If the path-hypotheses were not
``used'' anywhere in the match, Coq would notice this and
conclude that two invocations of the induction principle ought to be
judgmentally equal as soon as they have the same point-hypotheses,
even if their path-hypotheses differ, also leading to inconsistency.
(This was noticed by Bordg.)

The library currently includes higher inductive definitions of the
interval, the circle, suspensions, truncations, set-quotients, and
homotopy colimits such as coequalizers and pushouts, as well as the
flattening lemma, all from Chapter~6 of the HoTT book.  The library
also formalizes some basic synthetic homotopy theory from Chapter~8,
such as using the encode-decode method to prove that the fundamental
group of the circle is the integers.

The library contains some of the more experimental
higher inductive types from the HoTT book.  The cumulative hierarchy
of well-founded sets from Chapter~10 was formalized by
\citet{Ledent} and is now part of the library.  Two higher
inductive-inductive types from Chapter~11, the Cauchy reals
and the surreal numbers
are trickier. Even ordinary inductive-inductive
types~\citep{nordvallforsbergSetzer2012finIndind} are not supported in
Coq 8.5, but fortunately it is possible to simulate them using private
inductive types.  The surreals are thus in the library, while the Cauchy
reals have been formalized elsewhere~\citep{Gilbert}, as has another example of a higher inductive-inductive type, the
partiality monad~\citep{Altenkirch}. The formalization of the Cauchy reals and the partiality monad is based on an
experimental Coq implementation of inductive-inductive
types.\footnote{\cpporarxiv{Sozeau}{Sozeau}, \url{https://github.com/mattam82/coq/tree/IR}} We plan to include these developments into the library as soon as the support for inductive-inductive types is sufficiently mature.

Finally, we note that since private inductive types do not technically
use the path constructors, higher inductive types defined in this way
sometimes end up in a universe that is too low.  This can usually be
fixed with explicit universe annotations.


%
%


\section{\cpporarxiv{Modalities and Modules}{Modalities and modules}}\label{sec:modules}

Type-theoretic modalities, cf.\ Section~7.7 of the HoTT book and~\citep{RSS}, generalize the $n$-truncations in homotopy type theory. They were used in \emph{cohesive} homotopy type theory of \citet{schreibershulman2012}.
While we were able to formalize proofs and constructions about modalities in a straightforward manner, the overall
architecture of the definitions was quite tricky to implement.

Chapter~7 of the HoTT book contains a number of facts about truncations and
connectedness which actually hold (as
remarked therein) for any modality, with exactly the same proofs.
In the library we therefore formalized general facts about modalities,
following Chapter~7, and then instantiated them to truncations, which
we defined as particular instances of modalities.

A modality is an operator $\bigcirc$ which acts on types and satisfies a universal
property that quantifies over all types. One might expect the formalization of~$\bigcirc$ to be a
universe-polymorphic record type whose fields are the operator and its universal property.
This does not work however, because the fields would share a common universally quantified
universe index~$i$, and would thus express the wrong universal property. That is, we need
to express that $\bigcirc$ at level~$i$ has the universal propety with respect to every
level~$j$, not only~$i$.
We needed a construct like record types, but allowing each field to be
individually universe-polymorphic.

The solution was an undocumented feature of the Coq module system, which we briefly
review. Coq provides ML-style modules similar to those of OCaml, the programming language
used to implement Coq. Modules and type classes are similar, but the former are not
first-class objects and therefore cannot be used for formalization of structures. Instead,
modules serve a software-engineering purpose: they are used to organize source code into
units, and to control visibility of implementation details. By a fortunate design choice,
each entry in a module carries its own universal quantification over the universe levels.

The library formalizes a modality as a module type so that it can have the correct
universal property. However, we lose many of the benefits of record types and type
classes, as well as notational conveniences.

First, to pass a simple argument to a parametrized module (a functor in the
ML-terminology), we need to wrap it into a module, which considerably increases the
complexity of the code. Even worse, it prevents us from defining families of modalities,
such as $n$-truncations indexed by a truncation index~$n$. We used the standard trick of
passing from families to display maps, and included in the definition of modality an
additional field \lstinline|Modality : Type|, to be used as the domain of a display map
which encodes a family of modalities. For instance, truncations are implemented as a
single module which sets \lstinline|Modality| to \lstinline|trunc_index|.

Second, Coq is very strict about matching modules against module
types: it insists that the universe parameters and constraints match
the specified ones exactly. Thus, in the formalization we had to
battle Coq's universe heuristics and tightly control universes with
explicit annotations in both definitions and proofs. We persevered
with frequent use of \lstinline|Set Printing Universes| and
\lstinline|Show Universes|, but we cannot recommend manual treatment
of universe levels.

Third, Coq insists that every field in a module be fully polymorphic, while in several
places we wanted to express additional constraints on the universe levels. We found a way
around it which is far from ideal. It would be helpful if we could control universe
indices and constraints in modules more precisely, although we fear having to do even more
manual bookkeeping of universes. It would be interesting to attempt to formalize
modalities in Agda or Lean, although it is not clear how this could be accomplished, since
neither seem to support the kind of polymorphism that we needed.

Despite all these problems, the modalities code in the library is
quite usable; from the outside one rarely needs to worry about the
universe issues.  In particular, the $n$-truncation is defined as a
modality, and most of the basic theorems about truncation are obtained
by specialization from general theorems about modalities.  This seems
to work quite smoothly in the rest of the library, although the user
looking for a theorem about truncations has to know to check the
general theorems about modalities.


\section{\cpporarxiv{Technical Changes to Coq}{Technical changes to Coq}}\label{sec:coq-changes}

From the beginning, the development of the library has involved close collaboration with 
Coq developers, most notably Sozeau. A number of changes to the Coq
system were needed when we started, but today the library compiles with standard Coq~8.5.1 and later.
Our library has served as a testing platform for a number of new Coq features. Inefficiencies and bugs were
reported by us and quickly addressed by the Coq
developers. We mention several noteworthy changes to Coq:
\begin{itemize}
\item The inductive definition of the
identity type has a single constructor, and so Coq puts it in $\prop$, contrary to what is needed in homotopy type theory.
The \texttt{indices-matter} option of Coq, which was implemented already by Herbelin for the purposes of the
Foundations library~\citep{VV-foundations}, changes this behavior to the desired one.
\item We avoid the impredicative $\prop$ altogether and only use $\hprop$. An
element of $\hprop$ consists of a type together with a proof that the type has at most one element.
This small change makes the whole standard library unusable, and
many tactics stop working, too. The solution was rather drastic: we ripped out the standard library
and replaced it with a minimal core that is sufficient for the basic tactics to work. There is experimental work
that aims to disentangle the tactics and the standard libary,\footnote{Gallego Arias, \url{http://github.com/ejgallego/coq/tree/coqlib-cleanup-master}}
which we hope to use in the future.
\item The private inductive types are another experimental addition to Coq which allowed us to
  implement higher inductive types. This was already discussed in \S\ref{HIT}.
\end{itemize}

\section{\cpporarxiv{Software Engineering}{Software engineering}}\label{sec:software-engineering}

The collaborative development of the library was made possible by
using modern software engineering tools:
\begin{itemize}
\item We use GitHub as a platform for version control and discussion of the code. We
  have a strict ``two pairs of eyes'' policy according to which a code change may be
  accepted only after it has been reviewed by two other developers. This encourages good
  collaboration and the use of standardized naming schemes.
\item An extensive style guide facilitates collaboration and allows
  others to contribute and build on the library. We use the GitHub wiki to keep track of 
 documentation, and we automatically generate browsable and replayable literate code with Coqdoc
 and Proviola\footnote{\url{http://mws.cs.ru.nl/proviola/}} tools.
  Our library is also one
 of the test cases for the JavaScript interface for Coq.\footnote{\url{https://x80.org/rhino-hott/}}
\item We use Travis\footnote{\url{https://travis-ci.org/}} for
  continuous integration. It checks
  whether a proposed code change compiles, which serves as a very useful sanity check. Travis also allowed us to keep track
  of compilation time, which may be an issue with such a large library. We built tools that help identify causes
  of performance degradation. When the culprit was a Coq change, we reported it to Coq developers who quickly fixed the issue.
\item The installation procedure for the HoTT library is fairly complicated, as it requires a customized Coq installation.
  We provide automated scripts that help users with the installation. We also made the library available through the
  very successful OCaml Package Manager (OPAM).
\item The use of Proof General together with Emacs TAGS allows us to
  easily navigate the library and find definitions of
  terms; Company-coq\footnote{\url{https://github.com/cpitclaudel/company-coq}} provides even better facilities.
\end{itemize}


\section{\cpporarxiv{Related Work}{Related work}}\label{sec:related-work}

The HoTT library was initiated as an attempt to understand the contents of Voevodsky's
Foundations library~\citep{VV-foundations}. In the beginning we closely followed the order
of development of concepts in the Foundations library, but usually with our own proofs.

Nowadays the Foundations library has been incorporated 
into UniMath \citep{UniMath}. UniMath and HoTT still share many ideas about
how to formalize homotopy type theory, with a few differences: UniMath uses
the inconsistent assumption $\type\,{:}\,\type$ as a simplifying device, whereas we deal
with all the complexities of universe polymorphism; UniMath takes a conservative approach
with respect to advanced Coq technology, whereas our library actually inspired a
number of Coq features and serves as a testbed for new ones; finally, the HoTT library
uses higher inductive types, which are generally avoided by UniMath.

Various external developments have also used the HoTT library as a
base, including libraries for the interpretation of the database query language SQL
in homotopy theory~\citep{HoTTSQL}, and for monadic semantics of probabilistic computation with continuous
data types~\citep{FS}.

There are implementations of homotopy type theory in other proof assistants. The HoTT-Agda
library\footnote{\url{https://github.com/HoTT/HoTT-Agda}} was initially started as a
parallel development in Agda, but quickly took a different direction and experienced
extensive development of synthetic homotopy theory. The private inductive types were
inspired by Licata's trick, which was originally implemented in Agda.

A new formalization of homotopy type theory in the proof assistant Lean~\citep{lean} is
growing at an impressive rate. Lean’s type class system is a particularly useful feature. 

A related development is the prototype implementation of cubical type
theory~\citep{Cubical}, which includes a computational interpretation of univalence and (some)
higher inductive types. This improves on one of the limitations of our setting, the use of axioms for
univalence and function extensionality, which block computation in some proofs. 
We look forward to the integration of this technology in proof assistants.

\section{Conclusion}\label{sec:conclusion}

We have developed a large, well-designed, and well-docu\-mented library for homotopy type
theory, which formalizes a large portion of the HoTT book, including higher inductive
types, and employs universe polymorphism.
The library has successfully been used as a basis for several more specific
formalizations.  It serves as a testing ground for new Coq features,
as well as a foundation on which to experiment with formalizing new
ideas and applications of homotopy type theory.

\cpporarxiv{
  \balance
  \bibliographystyle{abbrvnat}
  \bibliography{hott}
}{
  \renewbibmacro{finentry}{%
  \iffieldequalstr{entrykey}{RSS}
   {\finentry\balance}
   {\finentry}}
  \newcommand{\balancekey}{RSS}
  \printbibliography  
}

\end{document}

